\documentstyle[preprint,aps]{revtex}
\begin{document}
\preprint{preprint CBPF NF-032/03, to appear in Phys. Lett. A, hep-th/0311091.}
\tightenlines
\title{GENERALIZED LADDER OPERATORS FOR THE DIRAC-COULOMB PROBLEM
VIA SUSY QM}
\author{
R. de Lima Rodrigues\footnote{Permanent address: Departamento de
Ci\^encias Exatas e da Natureza, Centro de Forma\c{c}\~ao de Professores, Universidade Federal de Campina Grande,
Cajazeiras - PB, 58.900-000 - Brazil.
E-mail to RLR is rafaelr@cbpf.br or
rafael@df.ufcg.edu.br}
\\ Centro Brasileiro de Pesquisas F\'\i sicas (CBPF)\\
Rua Dr. Xavier Sigaud, 150, CEP 22290-180, Rio de Janeiro, RJ, Brazil
}

\maketitle

\begin{abstract}

Supersymmetry and the shape invariance
condition in quantum mechanics are applied as an algebraic method to
solve the Dirac-Coulomb problem.
The ground state and the excited states are investigated using
new generalized ladder operators.
\end{abstract}

\vspace{1cm}

PACS numbers:  03.65.Fd, 03.65.Ge, 11.30.Pb

\vspace{1cm}

\pacs
\newpage

\section{Introduction}

The supersymmetry (SUSY)
algebra in quantum mechanics (QM) began with the work of Nicolai \cite{N}
and was elegantly formulated by Witten \cite{W81}. This approach has attracted interest and was applied to construct the spectral resolution of
solvable potentials in many areas of physics \cite{bjd}.
SUSY QM was also formulated
by Gendenshtein who used the shape invariance property \cite{Gend83}.
The hydrogen atom was studied via SUSY QM in the non-relativistic context
by Kostelecky and Nieto \cite{Nieto85}. They used the SUSY QM for
spectral resolution and also for calculating transition probabilities
for alkali-metal atoms. Also, Zhang {et al.} have considered interesting applications of a semi-unitary
formulation of SUSY QM \cite{zhang00}.

The Dirac-Coulomb problem is an  exactly solvable
problem in relativistic quantum mechanics and the solution  can be found
in  books on quantum mechanics, see for instance
\cite{BD}. The Dirac-Coulomb problem has also
been studied via SUSY QM
\cite{hughes85,Sukud85,rau,Richard86,Jarvis86,Fred88,Nogami93,kwato94,fred95}.
Our purpose in this paper is to obtain the complete energy spectrum
and the energy eigenfunctions of the Dirac-Coulomb problem using the shape
invariance and new generalized ladder operators.
The Lie algebra associated with these operators for the shape invariant potentials
has been presented by Fukui-Aizawa \cite{fukui93} and
Balantekin \cite{Bala98}. This formalism 
has been applied to the exactly solvable potentials
in non-relativistic quantum mechanics \cite{arma99,fakhri03}.

It is particularly simple to apply SUSY QM
to shape-invariant potentials
because their SUSY partners  are similar in shape and differ only in the
parameters that appear in them. More specifically, if $V_{-}(x;a_{1})$ is
any potential, adjusted to have zero ground state energy $E^{(0)}_{-} = 0,$
its SUSY partner
$V_{+}(x;a_{0})$ must satisfy the requirement
$
V_+ (x;a_0) = V_- (x;a_1) + R(a_1),
$
where $a_{0}$ is a set of parameters, $a_{1}$ a function of the
parameter $a_{0}$ and $R(a_{1})$ is independent of $x$.
In this case, one can determine the energy levels
for $V_+(x;a_0),$ which correspond to $E^{(n)}_-= \sum^{n}_{i=1}R(a_i).$
In this context, recently, some relativistic shape invariant potentials have been investigated
\cite{Inv01}.

The SUSY hierarchical prescription was
used by Sukumar to solve  the energy spectrum of the
Dirac-Coulomb problem \cite{Sukud85}. In this work the Fukui-Aizawa-Balantekin
\cite{fukui93,Bala98} approach  to the
Dirac-Coulomb problem is investigated using SUSY QM.

The paper is organized as follows. In 
section II we realize a graded Lie algebra structure in terms of the
4x4 matrix supercharges, analogous to Witten's SUSY algebra for
the Dirac radial equation associated with the hydrogen atom. In
section III, using the shape invariance condition we deduce new generalized ladder operators in
relativistic quantum mechanics, via supersymmetry, in order to build
up the energy eigenvalue and eigenfunctions of supersymmetric
partner potentials. In
section IV some concluding remarks are given.

\section{THE DIRAC-COULOMB PROBLEM AND SUSY}

In this section, we adopt the Sukumar's
approach to construct the
4x4 matrix supercharges.
The Dirac radial equation for the hydrogen atom can be written as

\begin{equation}
\label{E1}
\left(
\begin{array}{cc}
\frac{dG}{dr} & 0\\
0 & \frac{dF}{dr}\end{array}\right) + \frac{1}{r}
\left(
\begin{array}{cc}
k & -\gamma\\
\gamma & -k\end{array}\right)\left(\begin{array}{c}G \\ F\end{array}\right)=
\left(
\begin{array}{cc}
0 & \alpha_1\\
\alpha_2 & 0\end{array}\right)\left(\begin{array}{c}G \\ F\end{array}\right),
\end{equation}
where $k$ is an eigenvalue of the Dirac operator
$K=\beta(\vec\Sigma \cdot \vec L + {\bf 1}),
\quad \gamma= \frac{ze^2}{c\hbar}, \quad \alpha_1= m+E, \quad
\alpha_2= m-E, \quad \mid k\mid= j + \frac 12 \quad (k= \pm 1, \pm 2,
\pm 3, \cdots), \vec\Sigma= (\Sigma_1, \Sigma_2, \Sigma_3)$ are a set of the Pauli  matrices,
$\vec L$ is the angular momentum operator and {\bf 1} is the 2x2 unity matrix.
 Now, let 
 $D$ be the operator which diagonalizes the matrix
that appears in the interaction term

\begin{equation}
D^{-1}(k\sigma_3-i\gamma\sigma_2)D=s\sigma_3,
\end{equation}
where $s= (k^2-\gamma^2)^{1\over 2}$. Then we get

\begin{equation}
\label{E7}
D= (s + k){\bf 1} +\gamma\sigma_1.
\end{equation}
Thus, we obtain

\begin{mathletters}
\label{generallabel ab}
\begin{equation}
\left(\frac{k}{s}+\frac{m}{E}\right)\tilde{F}=
\left(\frac{d}{d\rho}+\frac{s}{\rho}
-\frac{\gamma}{s}\right)\tilde{G},
\label{mlett:10a}
\end{equation}

\begin{equation}
\left(\frac{k}{s}-\frac{m}{E}\right)\tilde{G}=
\left(-\frac{d}{d\rho}+\frac{s}{\rho}-\frac{\gamma}{s}\right)\tilde{F},
\label{mlett:10b}
\end{equation}
\end{mathletters}
where

\begin{equation}
\label{E11a}
\left(\begin{array}{c}\tilde{G}\\ \tilde{F}\end{array}\right)=
D\left(\begin{array}{c}G\\F\end{array}\right), \quad \rho= Er.
\end{equation}

The eigenvalue equations for $k= \mid k\mid$ and $k= -\mid k\mid,$
respectively, become

\begin{mathletters}
\label{generallabel c}
\begin{equation}
\left(\frac{\mid k\mid}{s}+\frac{m}{E}\right)\tilde{F_+}=
\left(\frac{d}{d\rho}+\frac{s}{\rho}-\frac{\gamma}{s}\right)\tilde{G_+},
\label{mlett:12a}
\end{equation}

\begin{equation}
\left(\frac{\mid k\mid}{s}-\frac{m}{E}\right)\tilde{G_+}=
\left(-\frac{d}{d\rho}+\frac{s}{\rho}-\frac{\gamma}{s}\right)\tilde{F_+},
\label{mlett:12b}
\end{equation}
\begin{equation}
\left(\frac{-\mid k\mid}{s}+\frac{m}{E}\right)\tilde{F_-}=
\left(\frac{d}{d\rho}+\frac{s}{\rho}-\frac{\gamma}{s}\right)\tilde{G_-},
\label{mlett:12c}
\end{equation}
\begin{equation}
\left(\frac{-\mid k\mid}{s}-\frac{m}{E}\right)\tilde{G_-}=
\left(-\frac{d}{d\rho}+\frac{s}{\rho}-\frac{\gamma}{s}\right)\tilde{F_-},
\label{mlett:12d}
\end{equation}
\end{mathletters}
where
$
\tilde{F_{\pm}}= \tilde{F}(\pm\mid k\mid)$ and  $\tilde{G_{\pm}}= \tilde{G}(\pm\mid k\mid).
$

Defining the intertwining operators as

\begin{equation}
\label{E18}
A_0^{(+)}= {\bf I}\frac{d}{d\rho}+\left(\frac{s}{\rho}-\frac{\gamma}{s}\right)\sigma_3,
\end{equation}

\begin{equation}
\label{E19}
A^{(-)}_0=\left(A_0^{(+)}\right)^{\dagger}= -{\bf I}\frac{d}{d\rho}+\left(\frac{s}{\rho}-\frac{\gamma}{s}\right)
\sigma_3
\end{equation}
where ${\bf I}$ denotes the 2x2 unit matrix and

\begin{equation}
\label{E20}
{\bf O}= \frac{m}{E}\sigma_1 +i\frac{\mid k\mid}{s}\sigma_2,
\end{equation}
we get

\begin{equation}
\label{E22}
A_0^{(+)}\left(\begin{array}{c}\tilde{G_+}\\ \tilde{F_+}\end{array}\right)=
{\bf O}\left(\begin{array}{c}\tilde{G_+}\\ \tilde{F_+}\end{array}\right),
\end{equation}

\begin{equation}
\label{E23}
A_0^{(-)}\left(\begin{array}{c}\tilde{F_-}\\ \tilde{G_-}\end{array}\right)=
-{\bf O}\left(\begin{array}{c}\tilde{F_-}\\ \tilde{G_-}\end{array}\right),
\end{equation}
where ${\bf I}$ denotes the 2x2 unit matrix.



Using the result
\begin{equation}
A_0^{-}=-\sigma_1 A_0^{+} \sigma_1,
\end{equation}
 we see that there exist the supersymmetric
partner eigenvalue equations:

\begin{equation}
\label{ea}
A_0^{(-)}A_0^{(+)}\left(\begin{array}{c}\tilde{G_+}\\
\tilde{F_+}\end{array}\right)= \left(1+\frac{\gamma^2}{s^2}-
\frac{m^2}{E^2}\right)\left(\begin{array}{c}\tilde{G_+}\\
\tilde{F_+}\end{array}\right)
\end{equation}
and

\begin{equation}
\label{E29}
A_0^{(+)}A_0^{(-)}\left(\begin{array}{c}\tilde{F_-}\\ \tilde{G_-}\end{array}
\right)= \left(1+\frac{\gamma^2}{s^2}-\frac{m^2}{E^2}\right)
\left(\begin{array}{c}\tilde{F_-}\\ \tilde{G_-}\end{array}\right).
\end{equation}

The mutually adjoint non-Hermitian supercharge operators for
Witten's model are given by

\begin{equation}
\label{E26} Q_{+} = A_0^{(+)}\sigma_-=\left(
\begin{array}{cc}
0 & A_0^{(+)} \\
0 & 0
\end{array}\right)_{4x4},\quad
Q_{-} = A_0^{(-)}\sigma_+= \left(
\begin{array}{cc}
0 & 0 \\
A_0^{(-)} & 0
\end{array}\right)_{4x4},
\end{equation}
so that the SUSY Hamiltonian $H$ takes the form
\begin{eqnarray}
\label{HS}
H = [Q_+,Q_- ]_+
&=& \left(
\begin{array}{cc}
H_{-} & 0 \\
0 & H_{+}
\end{array}\right),
\end{eqnarray}
and  satisfies
$[H, Q_{\pm}]_-=0.$

 The pair of SUSY hamiltonians is given by
\begin{equation}
H_{\pm}=-{\bf I}{d^2\over d\rho^2}+W^2\mp {{dW}\over d\rho}
\end{equation}
where the matrix superpotential is given by
$W(\rho)= \left(\frac{s}{\rho}-\frac{\gamma}{s}\right)\sigma_3.$

\section{SPECTRAL RESOLUTION VIA LADDER OPERATORS}

Let us now build up the energy eigenvalues  and eigenfunctions
of supersymmetric partner
potentials using the shape invariance condition and
the generalized ladder operators.
From the last section one obtains the following matrix forms of the pair
of SUSY potentials:

\begin{equation}
V_-(\rho, \lambda, s)=
\left(\begin{array}{cc}\frac{s(s-1)}{\rho^2}-\frac{2\gamma}{\rho}+\frac{\gamma^2}{s^2} & 0\\0 & \frac{s(s+1)}{\rho^2}-\frac{2\gamma}{\rho}+\frac{\gamma^2}{s^2}\end{array}\right),
\end{equation}
\begin{equation}
V_+(\rho, \lambda, s)=
\left(\begin{array}{cc}\frac{s(s+1)}{\rho^2}-\frac{2\gamma}{\rho}+\frac{\gamma^2}{s^2} & 0\\0 & \frac{s(s-1)}{\rho^2}-\frac{2\gamma}{\rho}+\frac{\gamma^2}{s^2}\end{array}\right).
\end{equation}
Although the SUSY partner potentials  $V_{(\pm)}$ are not shape
invariant, we can see that their components are:

\begin{equation}
\label{V22}
V_{(+)11}(\rho, \gamma, s)= V_{(-)11}(\rho, \gamma, s+1)-\frac{\gamma^2}{(s+1)^2}+\frac{\gamma^2}{s^2},
\end{equation}

\begin{equation}
\label{V23}
V_{(-)22}(\rho, \gamma, s)= V_{(+)22}(\rho, \gamma, s+1)-\frac{\gamma^2}{(s+1)^2}+\frac{\gamma^2}{s^2}.
\end{equation}

From (\ref{sia}), (\ref{V22}) and (\ref{V23}) one can written

\begin{equation}
R_{11}(a_1)= R_{22}(a_1)= -\frac{\gamma^2}{(s+1)^2}+\frac{\gamma^2}{s^2}= \frac{\gamma^2}{a_0^2}-\frac{\gamma^2}{a_1^2},
\end{equation}
where $a_0=s$ and $a_1=a_0+1,$
so that $a_i=a_0+i, R_{11}(a_i)= \frac{-\gamma^2}{(s+i)^2}+\frac{\gamma^2}{s^2}=
\frac{-\gamma^2}{(a_i)^2}+\frac{\gamma^2}{a_0^2}.$
Thus we get the following energy eigenvalues of $H_{(-)11}= H_{(+)22}:$

\begin{equation}
\label{E11}
E^{(n)}_{-11}= E^{(n)}_{+22}= \sum^{n}_{i=1}R_{11}(a_i)
=-\gamma^2\sum^{n}_{i=1}\left(\frac{1}{(a_i)^2}-\frac{1}{a_{i-1}^2}\right)= \frac{-\gamma^2}{(s+n)^2}+\frac{\gamma^2}{s^2}.
\end{equation}
 Comparing (\ref{E11}) with the eigenvalue
equation (\ref{E29}) we obtain the energy eigenvalues of the
hydrogen relativistic atom, viz.,

\begin{equation}
1+\frac{\gamma^2}{s^2}-\frac{m^2}{E^{(n)^2}}= \frac{\gamma^2}{s^2}
-\frac{\gamma^2}{(s+n)^2}, 
\end{equation}
providing

\begin{equation}
E^{(n)}= \sqrt{\frac{m^2}{1+\frac{\gamma^2}{(\sqrt{k^2 - \gamma^2}+n)^2}}}, \quad n= 0, 1, 2,\cdots,
\end{equation}
which is in agreement with the result obtained by Sukumar
using the SUSY Hamiltonian hierarchy method \cite{Sukud85}.

Note that the shape invariance condition is associated
with translation of the parameters $a$'s,
so that the Eq. (\ref{V22}) can be written in the following form

\begin{equation}
\label{sia}
A_{11}^{(-)}(a_0)A_{11}^{(+)}(a_0)=A_{11}^{(+)}(a_1)A_{11}^{(-)}(a_1)+
R_{11}(a_1),
\end{equation}
where
\begin{equation}
\label{A11}
A_{11}^{(\pm)}(s)= \pm\frac{d}{d\rho}+
\frac{s}{\rho}-\frac{\gamma}{s}, \quad a_0=s.
\end{equation}

Following Fukui-Aizawa-Balantekin approach \cite{fukui93,Bala98},
we obtain the following ladder operators

\begin{equation}
\label{B-}
B_-(s)= T^{\dagger}(s)A^-_{11}(s), \quad B_+(s)= B^{\dagger}_-(s),
\end{equation}
where $T(s)$ being a translation operator defined by

\begin{equation}
\label{T}
T(s)= e^{\frac{\partial}{\partial s}},
\end{equation}
with  $T^{\dagger}(s)=
e^{-\frac{\partial}{\partial s}}.$ In this case we have the identity
$R(a_n)=T(a_0)R(a_{n -1})T^{\dagger}(a_0)$  and the folowing 
algebra  

\begin{equation}
\label{aT}
[B-,B_+]=T(a_0)R(a_0)T^{\dagger}(a_0)=R(a_1), \quad R(a_n)B_+(a_0)=B_+(a_0)R(a_{n-1}),
\end{equation}
where the shape invariance provides us
the translations of the parameters $a_n,$ viz., $a_n=a_0+n,$
valid for any $n.$
Thus, it is easy to see that the operators $B_{\pm}(a_0)$ and $R(a_n)$ satisfy
the following commutation relations

\begin{equation}
\label{ga}
[H_{(-)11}, B_+^n]= (R(a_1)+R(a_2)+\cdots +R(a_n))B_+^n, 
\end{equation}
and

\begin{equation}
\label{ga2}
[H_{(-)11}, B_-^n]= -B_-^n(R(a_1)+R(a_2)+\cdots +R(a_n)), \quad n=1, 2, \cdots,
\end{equation}
with $H_{(-)11}=B_+B_-=A_{11}^{(+)}A_{11}^{(-)}.$
Consequently we see that the $\tilde{F}_-^{(n)},$ component eigenfunction of the $n$-th excited stated, is given by

\begin{equation}
\tilde{F}_-^{(n)} \propto B_+^n(s)\tilde{G}_-^{(0)}(\rho; s),
\quad n=1, 2, 3,
\cdots_.
\end{equation}

The ground state eigenfunction must be annihilated by $B_-(s),$
then

\begin{equation}
A^-(s)\tilde{G}_-^{(0)}(\rho; s)= 0,
\end{equation}
which leads us to the following physically acceptable solution

\begin{equation}
\tilde{G}_-^{(0)}(\rho; s)= N_G
\rho^s e^{-\frac{\gamma}{s}\rho},
\end{equation}
where $N_G$ being the
normalization constant.

>From $(\ref{B-})$ and $(\ref{T})$ we see that the raising operator may be written as

\begin{equation}
B_+(s)= \left(\frac{d}{d\rho}+\frac{s}{\rho}-
\frac{\gamma}{s}\right)e^{\frac{\partial}{\partial s}}.
\end{equation}
Consequently, for the first excited state one may write

\begin{eqnarray}
\tilde{F}^{(1)}_-(\rho; s)&&=B_+(s)\tilde{G}^{(0)}_-(\rho; s)
\nonumber\\
{}&&\propto\left(\frac{d}{d\rho}+\frac{s}{\rho}-
\frac{\gamma}{s}\right)e^{\frac{\partial}{\partial s}}
\rho^s e^{-\frac{\gamma}{s}\rho}
\nonumber\\
{}&&=
\left[(2s+1)\left(\frac{1}{\rho}-\frac{\gamma}{s(s+1)}\right)\right]
\rho^{s+1}e^{-\frac{\gamma\rho}{s+1}}.
\end{eqnarray}

Finally we would like to call attention to the fact that the above formalism may be applied to the exactly solvable potentials
in relativistic quantum mechanics \cite{Inv01,antonio03}.
Also, the applications of semi-unitary transformations to construct supersymmetric partner Hamiltonian in non-relativistic quantum mechanics \cite{zhang00} can be implemented for the 
Dirac-Coulomb problem.

\section{Conclusion}

In this paper we investigated the
Dirac-Coulomb problem via supersymmetry in quantum mechanics.
The shape invariant formalism for the
supersymmetric partners is applied to obtain the complete energy
spectrum and eigenfunctions of the Dirac-Coulomb problem.



Our approach uses the algebraic structure for shape invariant potential,
recently proposed by Fukui-Aizawa-Balantekin \cite{fukui93,Bala98} .
This approach is different from the SUSY Hamiltonian hierarchy method applied by Sukumar
\cite{Sukud85}.

\vspace{1.0cm}

\centerline{\bf ACKNOWLEDGMENTS}

\vspace{1.0cm}

The author is grateful to A. N. Vaidya, whose advises and
encouragement were precious.
RLR was supported in part by CNPq (Brazilian Research Agency).
He wishes to thank J. A. Helayel Neto for the kind of hospitality
at CBPF-MCT. The author wishes also to thank the staff of the
CBPF and DCEN-CFP-UFCG.

\end{document}